\documentclass[runningheads]{llncs}
\usepackage{graphicx}

\usepackage{tikz}
\usepackage{comment}
\usepackage{amsmath,amssymb} 
\usepackage{color}
\usepackage{kotex}
\usepackage{subfig}
\usepackage{cite}
\usepackage{lineno}

\usepackage[accsupp]{axessibility}  

\begin{document}
\pagestyle{headings}
\mainmatter
\def\ECCVSubNumber{7200}  

\title{Pixel-by-pixel Mean Opinion Score (pMOS)\\ for No-Reference Image Quality Assessment} 

\titlerunning{pMOS for NR-IQA} 
\authorrunning{W. Kim et al.} 
\author{Wook-Hyung Kim, Cheul-hee Hahm, Anant Baijal, Namuk Kim, Ilhyun Cho and Jayoon Koo}
\institute{AI · Big Data Lab., Advanced R\&D Group, Samsung Electronics, South Korea}

\maketitle

\begin{abstract}
Deep-learning based techniques have contributed to the remarkable progress in the field of automatic image quality assessment (IQA). Existing IQA methods are designed to measure the quality of an image in terms of Mean Opinion Score (MOS) at the image-level (i.e. the whole image) or at the patch-level (dividing the image into multiple units and measuring quality of each patch). Some applications may require assessing the quality at the pixel-level (i.e. MOS value for each pixel), however, this is not possible in case of existing techniques as the spatial information is lost owing to their network structures. This paper proposes an IQA algorithm that can measure the MOS at the pixel-level, in addition to the image-level MOS. The proposed algorithm consists of three core parts, namely: i) Local IQA; ii) Region of Interest (ROI) prediction; iii) High-level feature embedding. The Local IQA part outputs the MOS at the pixel-level, or pixel-by-pixel MOS - we term it ‘pMOS’. The ROI prediction part outputs weights that characterize the relative importance of region when calculating the image-level IQA. The high-level feature embedding part extracts high-level image features which are then embedded into the Local IQA part. In other words, the proposed algorithm yields three outputs: the pMOS which represents MOS for each pixel, the weights from the ROI indicating the relative importance of region, and finally the image-level MOS that is obtained by the weighted sum of pMOS and ROI values. The image-level MOS thus obtained by utilizing pMOS and ROI weights shows superior performance compared to the existing popular IQA techniques. In addition, visualization results indicate that predicted pMOS and ROI outputs are reasonably aligned with the general principles of the human visual system (HVS).

\end{abstract}

\section{Introduction}
Image Quality Assessment (IQA) finds application in various fields ranging from image compression to image restoration and image enhancement \cite{mier2021deep,li2021quality,liang2016comparison,zhang2021no,zhai2021perceptual,bhatt2021ssim}. IQA may be classified into subjective and objective types. In case of subjective IQA, humans from the general public or domain experts assess the quality of the image and their collective ratings yield a Mean Opinion Score (MOS). However, owing to human involvement, subjective IQA is time-consuming and expensive, making it difficult to apply in real-life and real-time applications. On the other hand, objective or automatic IQA, also known as computational IQA, is performed using machines and is an important topic of interest among researchers. Recently, deep-learning based IQA has been receiving increasing attention \cite{zhai2020perceptual,gu2021ntire, talebi2018nima, bosse2016deep, kang2014convolutional, bianco2018use} by the community. The objective IQA is further classified into two categories: i) Full-Reference IQA (FR-IQA)- e.g. SSIM \cite{wang2004image}, MS-SSIM \cite{wang2003multiscale}; ii) No-Reference IQA (NR-IQA)- e.g. BRISQUE \cite{mittal2012no}, NIQE \cite{mittal2012making}, CORNIA \cite{ye2012unsupervised}, PIQE \cite{venkatanath2015blind}.  

Among them, NR-IQA, also known as Blind IQA (BIQA), is the most challenging because it involves measuring the quality of an image without depending on a corresponding reference image. In addition, low-level computer vision applications including image restoration and image enhancement (e.g. super resolution, de-blurring, de-noising) also often do not have reference images (i.e. original distortion free images) available, hence NR-IQA, among other IQA types, alone can be useful for assessing the image’s quality to provide tailored restoration or enhancement as required. 

Some low-level computer vision applications may require assessing the quality at the pixel-level, however, existing NR-IQA methods assess the quality at either the image-level (single MOS value for the whole image) or at the patch-level (dividing the image into multiple patches and assessing the quality for each patch), rendering their architecture unsuitable for pixel-level IQA. 

To solve the aforementioned problem, we propose deep-learning based pixel-by-pixel IQA, abbreviated as ‘pIQA’, comprising of three parts: i) Local IQA, that outputs MOS for each pixel; ii) Region of Interest (ROI) prediction, that outputs the weights representing relative importance of each region; iii) High-level feature embedding, that outputs high level features that are then embedded into the Local IQA part. The key differentiated features of proposed method are as follows:

1)	The underlying neural network structure for the Local IQA part excludes any network layers that could lead to loss of information at the pixel-level such as pooling layers.

2)	The ROI part, used for identifying perceptually important regions in the image, is learnt in a completed unsupervised manner without depending on complex visual attention mechanisms, eye-tracking apparatus or labelled datasets.

3)	The overall image-level MOS is obtained by utilizing both pMOS and ROI (weights) values, and exceeds the performance of existing popular IQA methods.

\section{Related Work}
The fundamental goal of automatic IQA is designing a computational model that can objectively assess the quality of an image such that the objective score is as close as possible to the ground-truth IQA score obtained via subjective tests. Depending on the availability of the reference image (clean or distortion free original image) during assessment, it may be classified into FR-IQA and NR-IQA.

The FR-IQA methods focus on measuring similarity or dissimilarity between the original image and the distorted image. Classical methods include Peak Signal to Noise Ratio (PSNR), Structural Similarity Index Measure (SSIM) \cite{wang2004image} that measures similarity between underlying structures, Multi-Scale SSIM (MS-SSIM) \cite{wang2003multiscale} that measures SSIM at multiple scales, Feature Similarity Index Measure (FSIM) \cite{zhang2011fsim} that compares certain low-level features such as zero crossing or certain regions and the Gradient Similarity Metric (GSM) \cite{liu2011image} that assesses the quality based on similarity of gradients between the distorted image and the reference image. As these methods involve the use of hand-crafted features, their performance deteriorates depending on the complexity of the characteristics of the image being assessed. Deep-learning based FR-IQA methods have  emerged \cite{hou2014blind,zhang2015som} and methods utilizing local normalized multi-scale Difference of Gaussian (DOG) \cite{bosse2016deep} for extracting image features at various image scales in a more efficient manner have also been proposed.

NR-IQA methods, where reference image is unavailable during assessment, have also been developed. The traditional NR-IQA methods appear similar in structure to the traditional image processing and computer vision tasks wherein experts and engineers typically design hand-crated features to calculate the MOS for the image. Several NR-IQA methods involve the use of Natural Scene Statistics (NSS) \cite{srivastava2003advances} and are based on the premise that the statistical features of distorted image differ from those associated with natural or undistorted images. BIQI \cite{moorthy2010two} is one such method that assesses the quality of an image based on Distorted Image Statistics (DIS). Other related methods include DIIVINE \cite{moorthy2011blind}, SSEQ \cite{liu2014no} and BLIINDS-II \cite{saad2012blind}. BRISQUE \cite{mittal2012no} is also an NSS-derived IQA method but differs in that it utilizes locally normalized features when measuring the deviation of image’s statistics from the NSS. On the other hand, non-NSS based methods include CORNIA \cite{ye2012unsupervised} and HOSA \cite{xu2016blind} which are dictionary-based methods trained through unsupervised feature learning.

However, the performance of NR-IQA methods was found to be limited, mainly due to the use of hand-crafted features, motivating the research community to try deep-learning based techniques. NR-IQA feature extraction based on deep-learning was designed by Ghadiyaram and Bovik \cite{ghadiyaram2014blind}. Another method using two CNNs utilizing image and gradient map was proposed in \cite{yan2018two}. In addition, a technique called RankIQA, which learns mutual rank orders between images for the same distribution type, was proposed in \cite{liu2016learning}, and another method called BLINDER, which uses features derived from VGG-16 network, was proposed in \cite{gao2018blind}. Hallucinated-IQA \cite{lin2018hallucinated} made an effort to integrate the human vision system into the pipeline. Research efforts were also made to generate dataset for NR-IQA using the General Adversary Network (GAN) \cite{goodfellow2016nips}. Whereas HyperIQA proposes a multiscale feature fused hypernetwork architecture with a two-step quality prediction procedure involving a semantic feature learning step followed by content-based image quality prediction step \cite{su2020blindly}, MetaIQA proposes a meta-learning approach that first learns meta-knowledge using various distortions which then serves as a prior model for quickly fine-tuning the quality model to unknown distortions \cite{zhu2020metaiqa}. TRIQ was the first technique to borrow the Transformer structure from the IQA \cite{you2021transformer}. In addition, TranSLA \cite{zhu2021saliency} improve the performance through reinforcing feature aggregation by adding a Boosting Interaction Part (comprising of saliency and gradient maps) to the model structure of TRIQ. 

While various works have greatly contributed to the progress of IQA, no studies have been conducted to assess the image quality at the pixel-level. In this work, we proposed a deep-learning based NR-IQA technique called pixel-by-pixel IQA (pIQA) that is capable of assessing the image quality at the pixel-level in terms of proposed pixel-level MOS (pMOS), and the obtained pMOS values are then used in conjunction with the proposed ROI weights to derive the overall quality score at the image-level.

\section{Proposed Method}
\begin{figure}[t]
	\centering
	\subfloat[Main network]{\includegraphics[width=\linewidth]{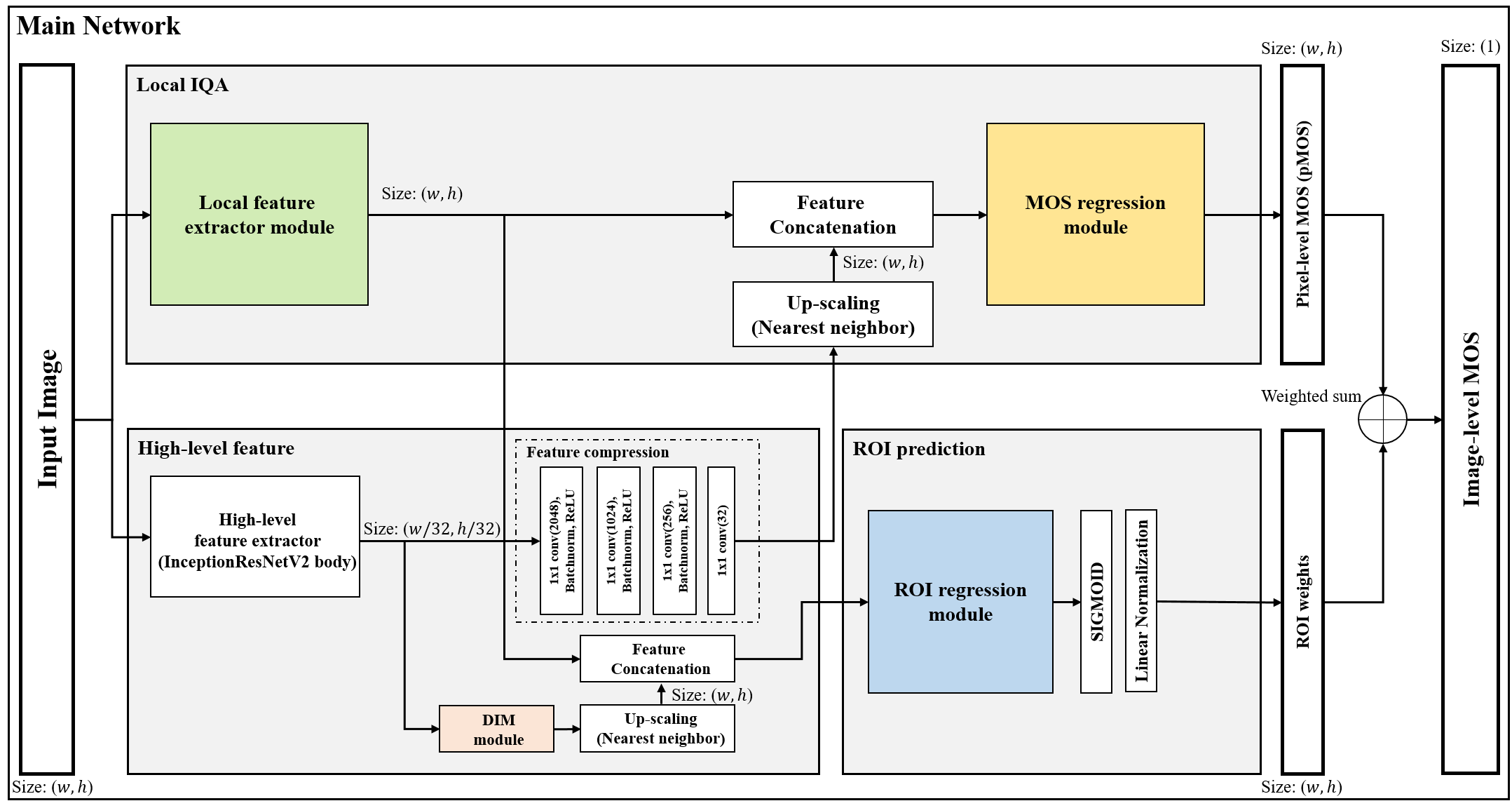} 
		\label{fig:architerture_a} }	
	\vfill
	\subfloat[Modules]{\includegraphics[width=\linewidth]{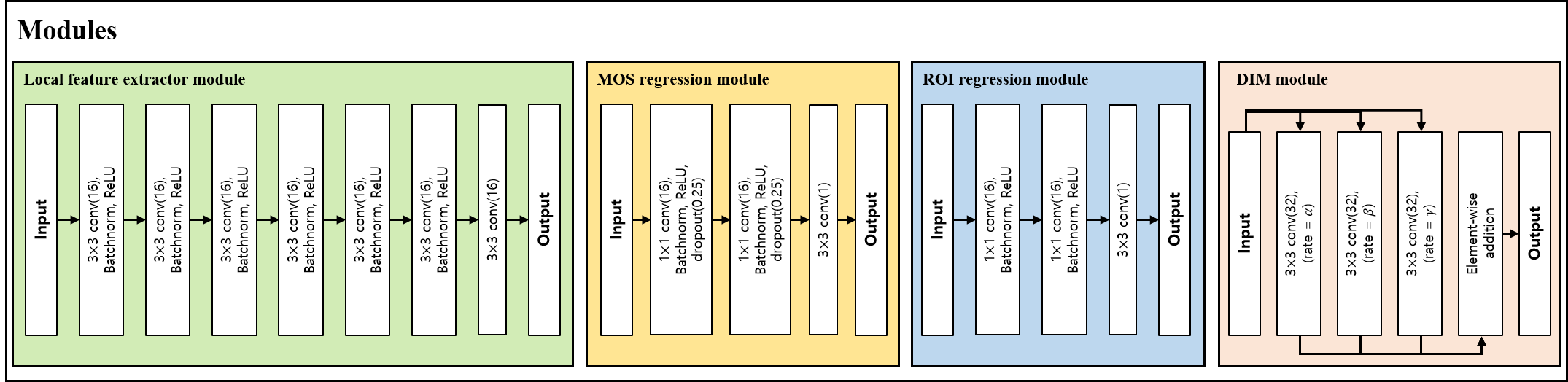} 
		\label{fig:architerture_b} }	
	\caption{Architecture of pIQA network. (a) Structure of the main network, (b) Structure of main network’s individual modules.}
	\label{fig:architerture}
\end{figure}

This section describes the proposed pIQA. The overall architecture of proposed method is shown in Fig. \ref{fig:architerture}. The proposed method consists of three parts: i) Local IQA; ii) ROI prediction; iii) High-level feature embedding. The Local IQA, is a neural network with a small receptive field and utilizes low-level features for outputting the MOS at the pixel-level i.e. MOS value for each pixel of the image (pMOS). The ROI prediction parts outputs the weights representing the relative importance of region, which are used when calculating the overall/image-level MOS. The High-level feature embedding part extracts high-level features that are used for both compensating the shortcomings of the Local IQA network when calculating the pMOS and as input to the ROI prediction part. The overall/image-level MOS is calculated as a weighted sum of pMOS values and ROI weights.

\subsection{Local IQA}
The Local IQA network is configured to utilize low-level image features and has a narrow receptive field in order to be able to measure the MOS at the pixel-level. As shown in Fig \ref{fig:architerture}, it consists of two modules, namely: Local feature extractor module and MOS regression module. Only convolution layer, batch normalization and ReLU are utilized in their construction. Moreover, only 10 convolution layers are used and the feature dimension is also kept minimal. Any layers such as strided convolution or pooling layers that lead to reduction in feature resolution are omitted from the network design in order to output MOS values which have same dimension with input image.
A total of seven 3$\times$3 2-D convolution layers towards the front (part of the local feature extractor module) extract the local features, and a total of three 1$\times$1 2D convolution layers towards the end (part of the MOS regression module) transform the extracted features into pMOS values. The regression module in the proposed structure plays a similar role as that of a fully connected layer present in the existing IQA networks.

\subsection{ROI prediction}
When a subjective evaluation is conducted, it is typical that not all regions of the image receive equal attention or consideration. For example, it is observed that during subjective evaluations, humans typically focus on the image regions rich in texture than flat regions and they focus more on the foreground than the background. Existing IQA methods make use of pooling layers and strided convolution layers to sample important regions while downsizing the image. However, in case of the proposed Local IQA part, since there is no such layer to sample important image regions, we design the proposed ROI prediction architecture that encapsulates the relative importance of image regions in terms of ROI weights. As shown in Fig. \ref{fig:architerture}, the ROI prediction part receives the extracted high-level and low-level features as input and outputs the ROI weights through the ROI regression module; the ROI regression module has the same structure as that of the MOS regression module but with the dropout layer removed.
\\~\\
\textbf{Linear normalization}
The proposed ROI prediction structure is similar to that of saliency prediction. Therefore, we added a normalization layer towards the end, as done in case of saliency prediction. We used linear normalization proposed in \cite{yang2019dilated} instead of the commonly used softmax normalization since it has been shown that linear normalization tends to reflect the probability distribution of human fixation more accurately than softmax normalization. Softmax normalization and linear normalization are shown in Eq (\ref{eq:softmax}) and Eq (\ref{eq:linear_norm}) respectively.

\begin{equation}
	\label{eq:softmax}
	s_{i,j} = \frac{exp(x_{i,j})}{\sum_{i=1}^{M}\sum_{j=1}^{N}exp(x_{i,j})}
\end{equation}

\begin{equation}
	\label{eq:linear_norm}
	r_{i,j} = \frac{x_{i,j}}{\sum_{i=1}^{M}\sum_{j=1}^{N}x_{i,j}}
\end{equation}			
where $\mathbf{x} = (x_{1,1}, ..., x_{M,N})$ is the set of unnormalizaed ROI prediction, $\mathbf{s} = (s_{1,1}, ..., s_{M,N})$ is the set of softmax normalizaed ROI prediction and $\mathbf{r} = (r_{1,1}, ...,\\ r_{M,N})$ is the set of linear normalizaed ROI prediction. $M$ and $N$ denote image width and height respectively.

\subsection{High-level feature embedding}
Since Local IQA has a narrow receptive field and a shallow network structure, adding high-level features can compensate for the shortcomings of the Local IQA. The high-level features used in our work are based on those used in \cite{hosu2020koniq} but differ in that we modify the underlying InceptionResNetV2 \cite{szegedy2016rethinking} used for extracting these features. In InceptionResNetV2, there is a layer with kernel-size of three but padding of 0.
Since the spatial resolution is reduced in absence of padding, we add padding where it is missing in order to preserve the spatial information of the features. For example, the 2D convolution layer and max-pooling layer of the ‘Reduction-A block’ in InceptionResNetV2 have a kernel size of 3 and a padding size of 0. In this case, however, the resolution is reduced by 2 pixels horizontally and vertically. Hence, the padding of size 1 was added to prevent the loss of spatial information. The network body before the global average pooling of the our modified InceptionResNetV2 was then used as a backbone network. Since the feature resolution extracted by the modified backbone network is 1/32 times the image size, the nearest neighbor upscaling is performed 32 times to match the local feature resolution with the size of the image. In addition, since the number of channels of the extracted high-level features is very large, the channels were compressed using a 1$\times$1 convolution layer before embedding the features to the local IQA. The features extracted from the Local feature extraction module and the High-level feature embedding are then concatenated inside the Local IQA part and passed through the MOS regression module to output the pMOS.
\\~\\
\textbf{Dilated inception module(DIM)}
DIM is a low-complexity method designed for predicting saliency utilizing a very wide receptive field \cite{yang2019dilated}. The DIM is added before the high-level feature are provided as input to the ROI regression module. In this case, the receptive field almost covers the entire image, so if there is a center-bias, it has the effect of being learnt implicitly. The high-level features after passing through the DIM are up-scaled and then concatenated with the local features before being provided as input to the ROI regression module.

\subsection{Loss fuction}
\begin{figure}[t] 
	\centering	
	\subfloat[KonIQ-10k]{\includegraphics[width=0.4\linewidth]{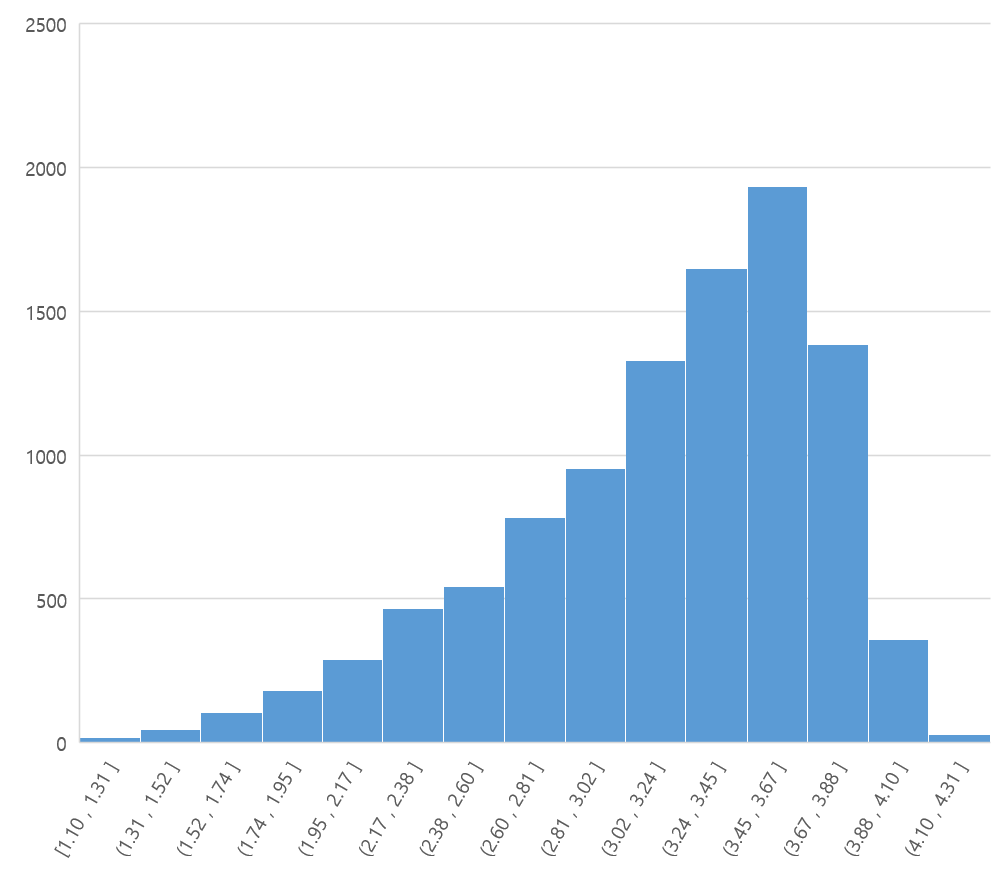} 
		\label{fig:mos_koniq} }	\hfil	
	\subfloat[LIVE Challenge]{\includegraphics[width=0.40\linewidth]{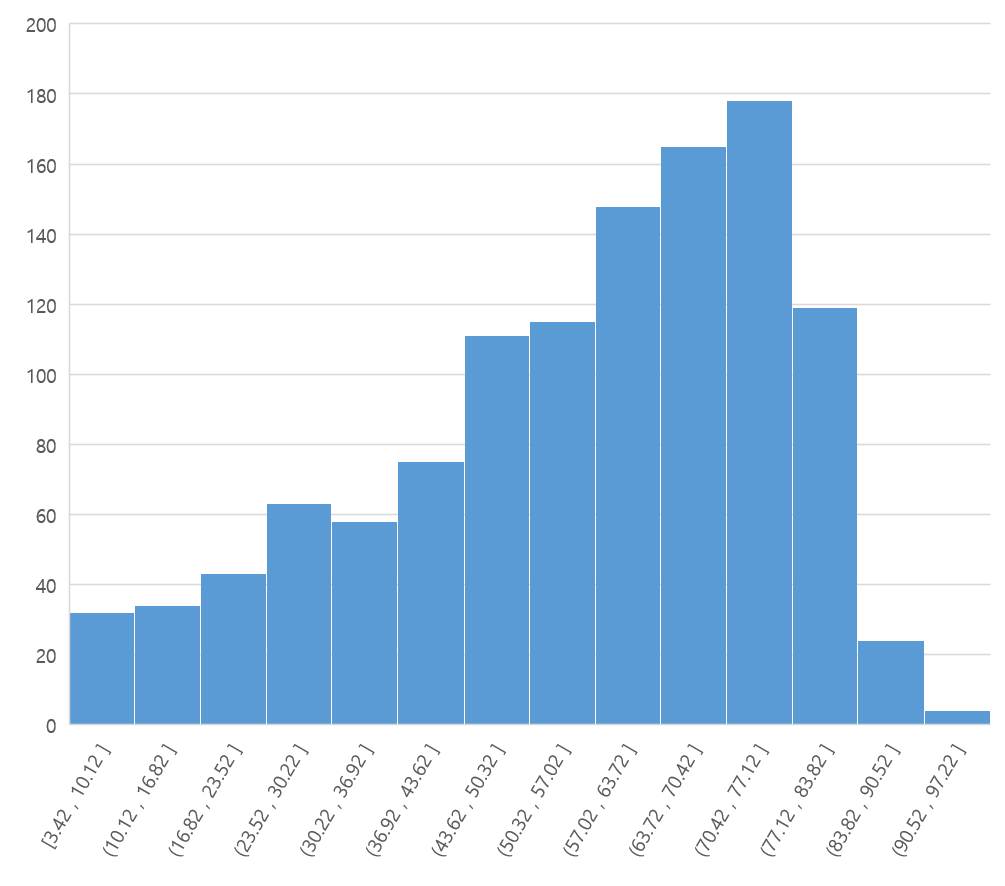} 
		\label{fig:mos_livec} }	
	\caption{MOS distribution of KonIQ-10k and LIVE Challenge dataset. The $x$-axis represents the MOS range and the $y$-axis represents the number of image samples.}
	\label{fig:mos_dist}
\end{figure}
The proposed network makes use of $L1$ loss during training.
\begin{equation}
\begin{split}	
	\label{eq:basic_loss}
	L(P, G) &= |P - G| \\
			&= |\sum_{i=1}^{M}\sum_{j=1}^{N}p_{i,j}\cdot r_{i,j} - G|
\end{split}
\end{equation}
$P$ and $G$ represent the predicted MOS and ground-truth MOS at the image-level, respectively, and $\mathbf{p} = (p_{1,1}, ..., p_{M,N})$ represents the set of predicted pMOS. Through this equation, ROI is learned in an unsupervised manner without depending on any ROI ground-truth.
\\~\\
\textbf{MOS mean shifted loss}
When the ROI network is learned via the Eq. (\ref{eq:basic_loss}), the $p_{i, j}$ (=pMOS) acts as the learning weight. As a result, when the MOS value is high, the ROI network inadvertently has high learning weight, and when the MOS value is low, it has a low learning weight. For example, when the MOS range is between 1 and 5, the learning speed of the ROI network can vary as much as 5 times. However, since this is not what we intended, we need to correct the learning weight.

Looking at the MOS distribution in Fig. \ref{fig:mos_dist}, the probability is highest towards the center and the lowest at both ends of the MOS range. Therefore, from an entropy perspective, both ends have more information than the distribution at the center of the MOS. If the mean value of $p_{i,j}$ is moved to 0 during the learning phase, the learning weight can be lowered towards the center of the MOS whereas a higher learning weight can be assigned towards the both ends, as a result a higher learning weight is assigned to the part of the distribution containing more important information. The final loss after MOS mean-shift is as follows:

\begin{equation}
	\begin{split}	
		\label{eq:meanshift_loss}
		L_{ms}(P_{ms}, G) &= |P_{ms} - G| \\
		&= |\sum_{i=1}^{M}\sum_{j=1}^{N}(p_{i,j}-\overline{p})\cdot r_{i,j} - G|,\\
		\overline{p} =&\frac{1}{MN}\sum_{i=1}^{M}\sum_{j=1}^{N}(p_{i,j})
	\end{split}
\end{equation}

\section{Experimental Results}
\subsection{Dataset}
IQA datasets can be largely divided into two types \cite{yang2019survey}. The first type is the synthetic dataset which is created by applying various types of distortions to a small number of reference images. The second type is called authentic or realistic dataset that contains images having authentic distortions. However, synthetic dataset lacks the diversity of images required to learn our ROI prediction part as the number of reference images is too small to be meaningful. Therefore, we conduct experiments utilizing KonIQ-10k \cite{hosu2020koniq} and LIVE Challenge \cite{ghadiyaram2015massive} dataset (LIVEC), two popular large-scale public datasets containing diverse images. 

The KonIQ-10K consists of a total of 10,073 images, each having a dimension of 1024$\times$768. These images were sampled from YFCC100m considering the distribution of contrast, color and sharpness. LIVEC consists of a total of 1,162 images each having a dimension of 500$\times$500. The images were captured in-the-wild by different photographers with a variety of camera devices and were evaluated by more than 8,100 subjects.

\subsection{Implementation details}
We use the Nvidia Tesla V100 GPU and the Pytorch library for our experimental environment. We make use of Adam optimizer and a total of 90 epoch trains, dividing them into three epoch trains of 30 each, with respective learning rates of $10^{-4}$, $5 \times 10^{-4}$ and $10^{-5}$. A batch size of 48 is used for KonIQ and 36 for LIVEC, and KonIQ is downsized by $\frac{1}{2}$ to conduct the experiment on images having size of 512$\times$ 384 each. In addition, random 180 degree rotations and random image flips are performed for data augmentation purpose. Like other state-of-the-art methods \cite{su2020blindly, wu2020end, zhu2020metaiqa, zhu2021saliency}, we split the dataset randomly- 80\% is used for training and the remaining 20\% is used for testing purposes. The parametric values for DIM (Fig. \ref{fig:architerture_b}) are set as $\alpha$=2, $\beta$=4 and $\gamma$=8. The ground-truth MOS values for LIVEC were shifted and scaled such that the the overall mean and variance of LIVEC become the same as that of KonIQ.

\subsection{Evaluation results and discussion}
We evaluate the performance of our proposed method based on Pearson’s Linear Correlation Coefficient (PLCC) and Spearman’s Rank Order Correlation Coefficient (SRCC), which are commonly used metrics for performance evaluation of IQA. Whereas PLCC measures the accuracy of the prediction, SRCC measures the monotonicity of the prediction. The values for both PLCC and SRCC range between -1 and 1 (higher value represents better the performance). Addidionaly, Root Mean Square Error (RMSE) results are also presented in Table \ref{table:compare}.
\\~\\
\textbf{A. Comparison with the State-of-the-art}
\begin{table}[t]
	\centering
	\begin{tabular}{l|ccc|ccc}
		\hline \hline
		\multicolumn{1}{c|}{}       & \multicolumn{3}{c|}{KonIQ-10k}                   & \multicolumn{3}{c}{LIVEC}                        \\ \hline
		\multicolumn{1}{c|}{Method} & PLCC           & SRCC           & RMSE           & PLCC           & SRCC           & RMSE           \\ \hline
		GraphIQA                    & 0.922          & 0.907          & -              & 0.886          & 0.863          & -              \\
		HyperIQA                    & 0.917          & 0.906          & -              & 0.882          & 0.859          & -              \\
		TRIQ                        & 0.923          & 0.902          & 0.225          & 0.878          & 0.849          & 0.392          \\
		TranSLA                     & 0.931          & 0.915          & 0.206          & 0.900          & 0.873          & 0.359          \\
		Proposed method             & \textbf{0.943} & \textbf{0.925} & \textbf{0.191} & \textbf{0.914} & \textbf{0.900} & 0.231 \\ \hline \hline
	\end{tabular}
	\caption{Comparative results.}
	\label{table:compare}	
\end{table}
We compared the performance of our method with other popular IQA methods. The comparative results are summarized in Table \ref{table:compare}, with the best-performance indicated in bold. As can be seen from Table \ref{table:compare}, our method achieves the highest PLCC and SRCC values on both KonIQ and LIVEC datasets. In particular, while RMSE shows a significant decrease of 7.3\% compared to existing techniques for KonIQ, we cannot provide the comparison for LIVEC as information on the MOS scale used by the previous works is unavailable.
\\~\\
\textbf{B. Ablation study}
\begin{table}[t]
	\centering
\begin{tabular}{l|ccc|ccc}
	\hline
	\multicolumn{1}{c|}{}                 & \multicolumn{3}{c|}{KonIQ-10k}                   & \multicolumn{3}{c}{LIVEC}                        \\ \hline
	\multicolumn{1}{c|}{Components}       & PLCC           & SRCC           & RMSE           & PLCC           & SRCC           & RMSE           \\ \hline\hline
	Local IQA                             & 0.697          & 0.664          & 0.436          & 0.595          & 0.628          & 0.522          \\
	Local IQA + ROI                       & 0.839          & 0.812          & 0.367          & 0.762          & 0.752          & 0.379          \\
	Local IQA + ROI + High-level featrues & \textbf{0.943} & \textbf{0.925} & \textbf{0.191} & \textbf{0.914} & \textbf{0.900} & \textbf{0.231} \\ \hline\hline
\end{tabular}
	\caption{Ablation study of proposed method.}
	\label{table:ablation}	
\end{table}
An ablation study was conducted to investigate the efficacy of the individual parts of the proposed method. Table \ref{table:ablation} shows the results obtained as each part is added one after the other. The network, when consisting of Local IQA alone, shows poor performance as it has learned to predict the MOS by assigning the same weights to all regions in an image. However, when the ROI prediction part is added, it can be seen that the performance is dramatically improved as the MOS is calculated through a weighted sum wherein more weight is assigned to perceptually more important regions of the image. Finally, when High-level feature embedding part is also added, the performance is again improved owing to the inclusion of high-level features which could not be captured by the Local IQA network alone.
\begin{figure} 
	\centering
	\captionsetup[subfigure]{justification=centering}	
	\subfloat[Image\label{fig:visual_a}]{\includegraphics[width=0.22\linewidth]{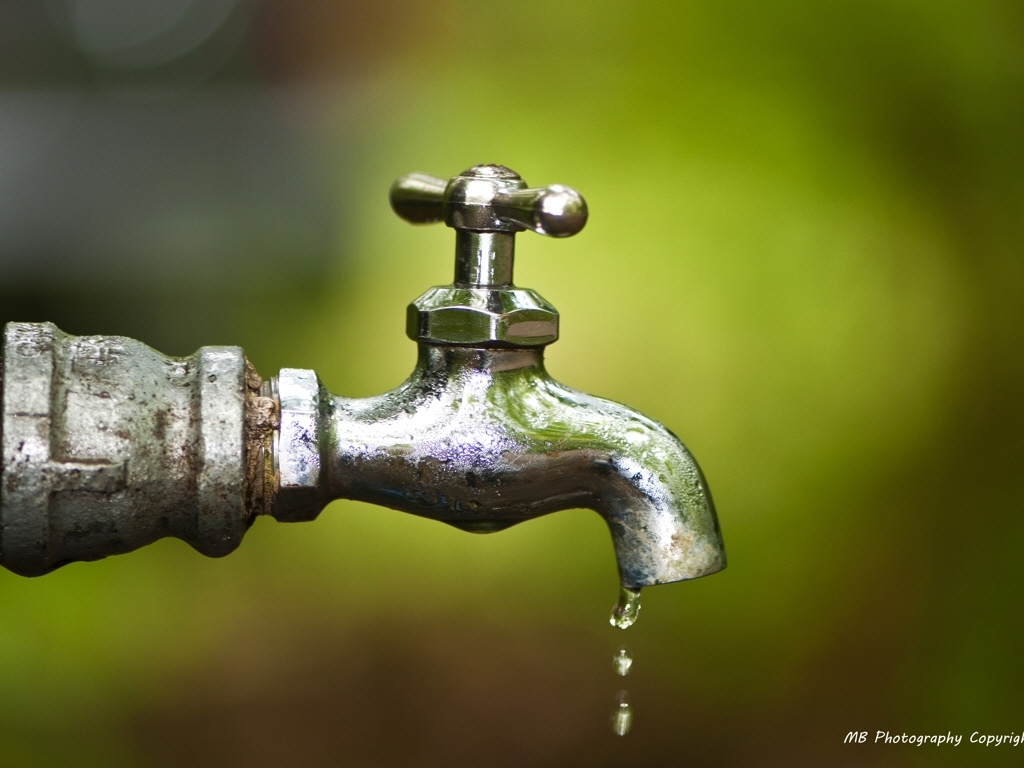}}	\hspace{0.05\linewidth}
	\subfloat[pMOS]{\includegraphics[width=0.22\linewidth]{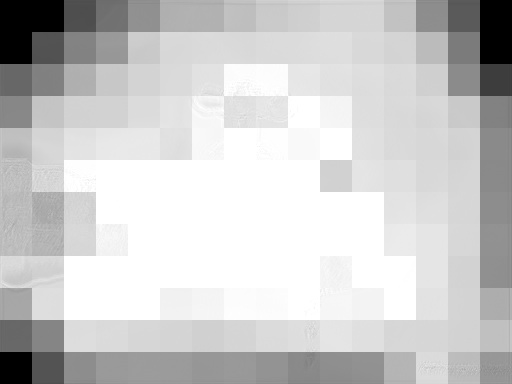}}	\hspace{0.05\linewidth}
	\subfloat[ROI]{\includegraphics[width=0.22\linewidth]{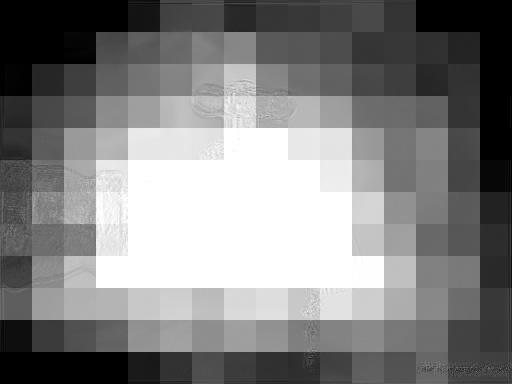}} 
	\vfill \hspace{0.22\linewidth}\hspace{0.05\linewidth}
	\subfloat[pMOS\\(local only)]{\includegraphics[width=0.22\linewidth]{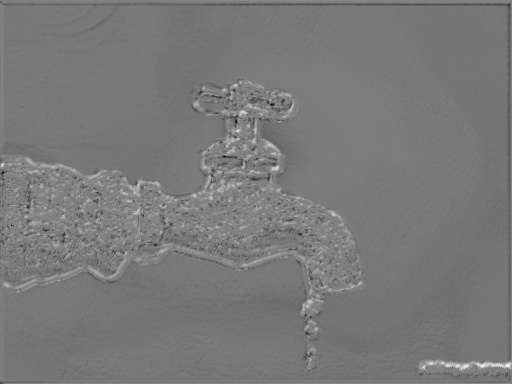}}\hspace{0.05\linewidth}
	\subfloat[ROI\\(local only)]{\includegraphics[width=0.22\linewidth]{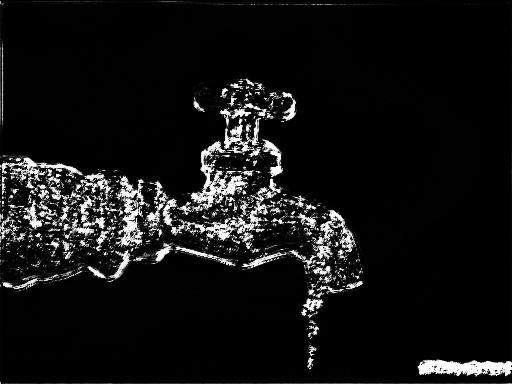}}

	\subfloat[Image]{\includegraphics[width=0.22\linewidth]{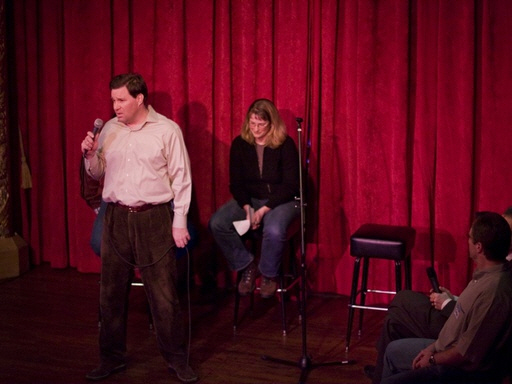}}\hspace{0.05\linewidth}	
	\subfloat[pMOS]{\includegraphics[width=0.22\linewidth]{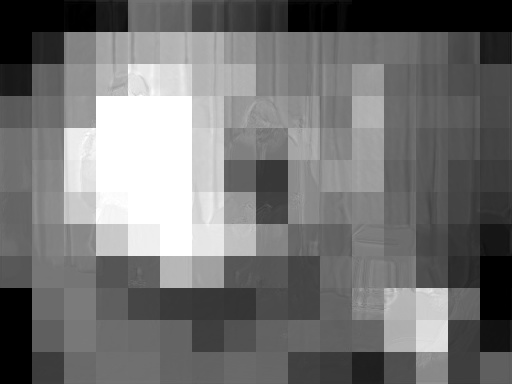}}\hspace{0.05\linewidth}
	\subfloat[ROI]{\includegraphics[width=0.22\linewidth]{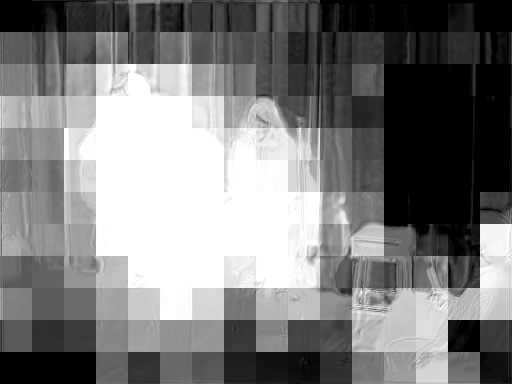}}	
	\vfill \hspace{0.22\linewidth}\hspace{0.05\linewidth}
	\subfloat[pMOS\\(local only)]{\includegraphics[width=0.22\linewidth]{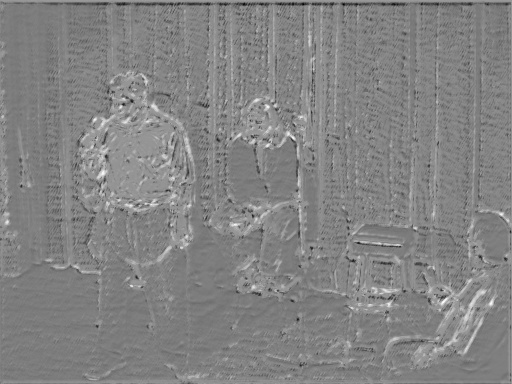}}\hspace{0.05\linewidth}
	\subfloat[ROI\\(local only)]{\includegraphics[width=0.22\linewidth]{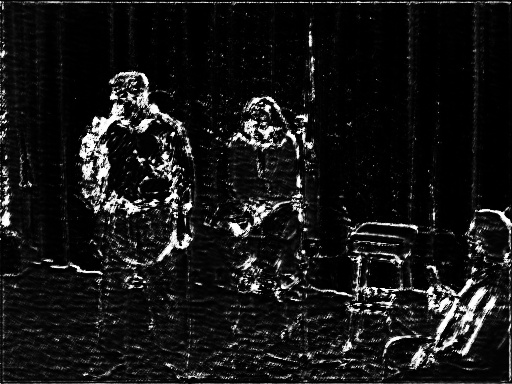}}
	
	\subfloat[Image]{\includegraphics[width=0.22\linewidth]{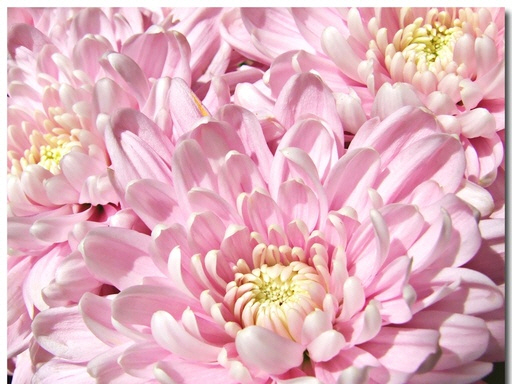}}\hspace{0.05\linewidth}
	\subfloat[pMOS]{\includegraphics[width=0.22\linewidth]{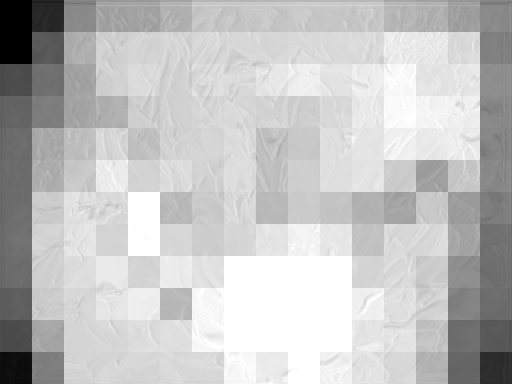}}\hspace{0.05\linewidth}
	\subfloat[ROI]{\includegraphics[width=0.22\linewidth]{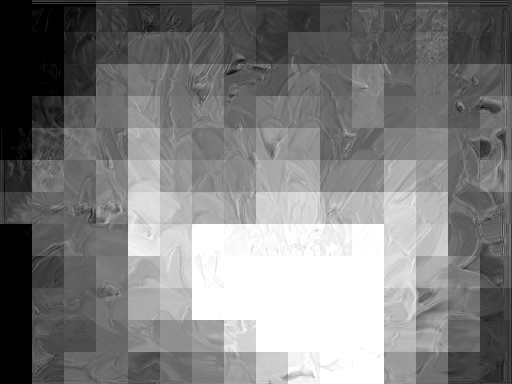}}	
	\vfill \hspace{0.22\linewidth}\hspace{0.05\linewidth}
	\subfloat[pMOS\\(local only)]{\includegraphics[width=0.22\linewidth]{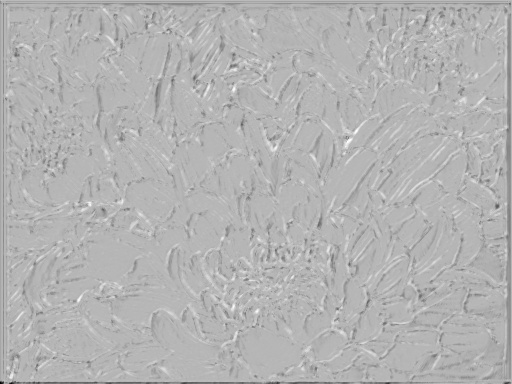}}\hspace{0.05\linewidth}
	\subfloat[ROI\\(local only)]{\includegraphics[width=0.22\linewidth]{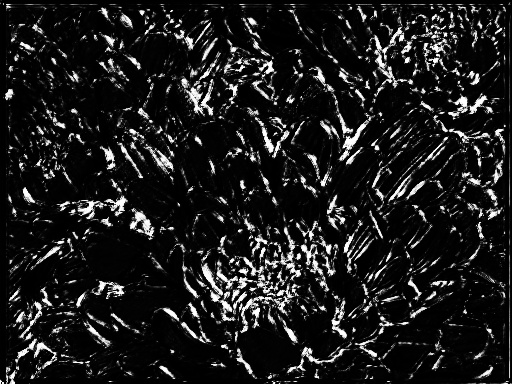}}
	\caption{Visualization results of the proposed method. The ‘local only’ excludes the usage of high-level features from the proposed method.}
	\label{fig:visual}
\end{figure}
\\~\\
\textbf{C. Visulaization}
Fig. \ref{fig:visual} shows the visualization results of the proposed method. The resultant image is displayed for two cases: when the network includes the high-level features (i.e. both local features and high-level features are used) and when the excludes the high-level features (i.e. only local features are used). In Fig \ref{fig:visual}, (a), (f) and (k) show the respective original images used as input to our method, (b), (g) and (l) show the respective pMOS, (c), (h) and (m) show the respective ROI prediction, (d), (i) and (n) show the respective pMOS when using the local-features alone, and (e), (j) and (o) show the respective ROI prediction results when using the local-features alone. As we can see from Fig. \ref{fig:visual}, the network including both high-level and local features appear to calculate pMOS and ROI by considering high-level concepts such as objects (for example people and flowers), in addition to low-level features such as edges. Especially for images such as Fig. \ref{fig:visual} (a) that contain a distinct foreground object and a background, we can see that pMOS values tend to be higher for the foreground pixels than the background pixels. Similarly, ROI shows higher weight values for the in-focus pixels compared to the out-of-focus pixels.
On the other hand, the network relying upon local feature alone appears to calculate pMOS and ROI mainly by analyzing the low-level features such as edges and texture. However, the pMOS values calculated using only the local-features shown a better pixel-level alignment due to the associated low-receptive field.

\section{Conclusion}
We propose a new IQA method called the pixel-by-pixel IQA (pIQA) that can calculate MOS values at the pixel-level (pMOS) in addition to the image-level, unlike existing IQA techniques that can calculate MOS value at the image-level alone. Moreover, our image-level MOS predicted through a weighted sum of pMOS and ROI achieves superior results (in terms of PLCC, SRCC and RMSE) when compared with existing popular methods. In addition, visualization results indicate that ROI output is reasonable and largely aligned with our general expectations in that the weights are found to be typically high for pixels containing objects, foregrounds and edges. Moreover, the nature of pMOS values resmbles that of MOS values in that both tend to have relatively higher values for in-focus or sharp regions than out-of-focus or flat regions.
In the future, we would like to focus on designing a lighter and more accurate network architecture in order to further improve its potential applicability to other low-level vision applications.

\clearpage
%
%
\bibliographystyle{splncs04}
\bibliography{egbib}
\clearpage

\section{Supplement material}
\begin{figure}[]
	\centering
	\subfloat[Mean ROI of KonIQ dataset with DIM]{\includegraphics[width=0.3\linewidth]{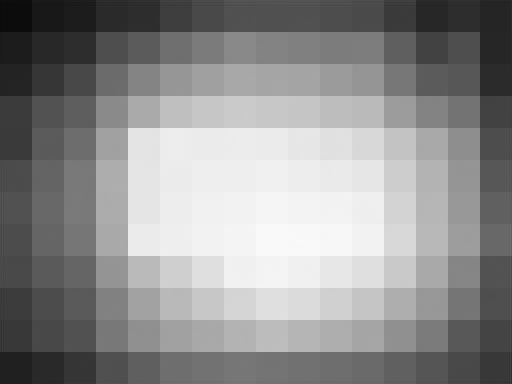}}
	\hspace{0.1\linewidth}
	\subfloat[Mean ROI of KonIQ dataset without DIM]{\includegraphics[width=0.3\linewidth]{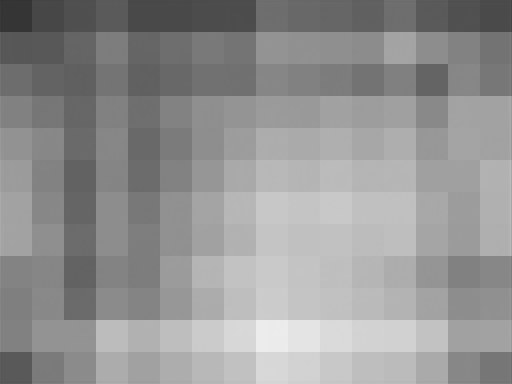}}
	\vfill
	\subfloat[Mean ROI of LIVEC dataset with DIM]{\includegraphics[width=0.3\linewidth]{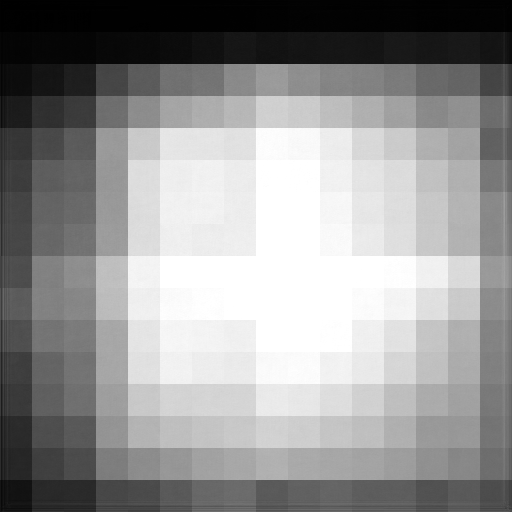}}
	\hspace{0.1\linewidth}
	\subfloat[Mean ROI of LIVEC dataset without DIM]{\includegraphics[width=0.3\linewidth]{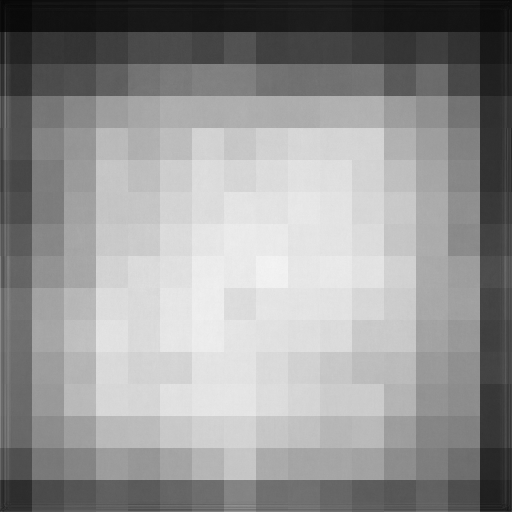}}
	\caption{Visual results obtained for with and without DIM cases. The use of DIM results in center-bias phenomenon.}
	\label{fig:dim}
\end{figure}

\begin{table}[]
	\centering
	\begin{tabular}{l|ccc|ccc}
		\hline \hline
		\multicolumn{1}{c|}{}       & \multicolumn{3}{c|}{KonIQ-10k}                   & \multicolumn{3}{c}{LIVEC}                        \\ \hline
		\multicolumn{1}{c|}{Method} & PLCC           & SRCC           & RMSE           & PLCC           & SRCC           & RMSE           \\ \hline
		Without MS loss             & 0.871          & 0.862          & 2.551          & 0.795          & 0.801          & 2.603          \\
		Without DIM module          & 0.942          & \textbf{0.925} & 0.193          & \textbf{0.916} & \textbf{0.900} & \textbf{0.228} \\
		With softmax normalization  & 0.934          & 0.913          & 0.206          & 0.902          & 0.880          & 0.251          \\
		Proposed method             & \textbf{0.943} & \textbf{0.925} & \textbf{0.191} & 0.914          & \textbf{0.900} & 0.231          \\ \hline \hline
	\end{tabular}
	\caption{Effect of various techniques.}
	\label{table:supplement}
\end{table}
~\\
\textbf{Effect of MOS mean-shift loss (MS loss)}
Table \ref{table:supplement} shows the learning results obtained for the case when MS loss is excluded (‘without MS Loss’). These results are obtained when learning is performed based on Eq. (3) instead of Eq. (4).  As can be seen from these results, the performance decreases significantly when MS loss is not used; RMSE values are especially high indicating that learning has not been done properly.  Visualizations shown in Fig. \ref{fig:dim} indicate that the ROI prediction for ‘without MS loss’ case is not as accurate when compared to that for the ‘with MS loss’ case. 
\begin{figure}[t] 
	\centering
	\captionsetup[subfigure]{justification=centering}	
	\subfloat[Image\label{fig:visual_a}]{\includegraphics[width=0.27\linewidth]{./figures/focus/high/ori/6118735934.jpg}}	\hspace{0.05\linewidth}
	\subfloat[ROI W/O MS loss]{\includegraphics[width=0.27\linewidth]{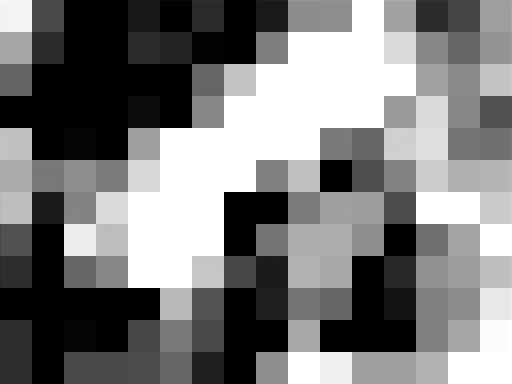}}	\hspace{0.05\linewidth}
	\subfloat[ROI with MS loss]{\includegraphics[width=0.27\linewidth]{./figures/focus/high/roi/6118735934.jpg}} 
	\vfill \hspace{0.25\linewidth}
	
	\subfloat[Image\label{fig:visual_a}]{\includegraphics[width=0.27\linewidth]{./figures/visual/high/ori/2365109175.jpg}}	\hspace{0.05\linewidth}
	\subfloat[ROI W/O MS loss]{\includegraphics[width=0.27\linewidth]{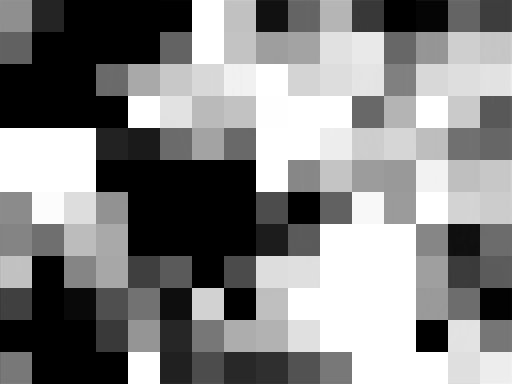}}	\hspace{0.05\linewidth}
	\subfloat[ROI with MS loss]{\includegraphics[width=0.27\linewidth]{./figures/visual/high/roi/2365109175.jpg}} 
	\vfill \hspace{0.25\linewidth}
	
	\subfloat[Image\label{fig:visual_a}]{\includegraphics[width=0.27\linewidth]{./figures/visual/high/ori/628981585.jpg}}	\hspace{0.05\linewidth}
	\subfloat[ROI W/O MS loss]{\includegraphics[width=0.27\linewidth]{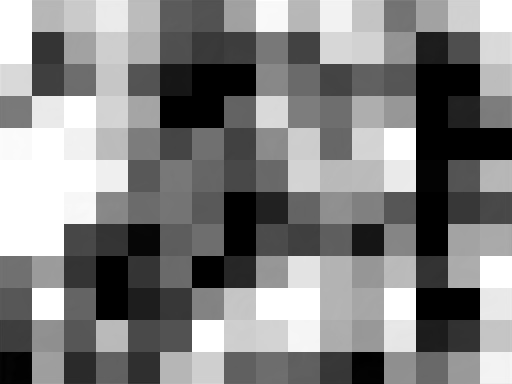}}	\hspace{0.05\linewidth}
	\subfloat[ROI with MS loss]{\includegraphics[width=0.27\linewidth]{./figures/visual/high/roi/628981585.jpg}} 
	\vfill \hspace{0.25\linewidth}
	\caption{Visualization results for the effect of MS loss.}
	\label{fig:ms_loss}
\end{figure}
\\~\\
\textbf{Effect of Dilated inception module (DIM)}
In order to test the effectiveness of DIM, we re-performed the experiment by replacing DIM's dilated 2D convolution layer with a normal 2D convolution layer instead. Fig. \ref{fig:dim} shows the visual results which are obtained by averaging the ROI prediction for all the test images. (a) and (b) are the results of 'with DIM' and 'without DIM' cases for the KonIQ dataset, respectively, and (c) and (d) are the results of 'with DIM' and 'without DIM' cases for the LIVEC dataset, respectively. We observe that whereas the center-bias effect is relatively stronger for the KonIQ dataset for the ‘with DIM’ case, it is relatively weaker for LIVEC dataset owing to few image samples (225 test images) although the center-bias effect is found to be more pronounced than for the ‘without DIM’ case. However, we observe little difference in the PLCC, SRCC, and RMSE results inferred from Table \ref{table:supplement}- this may be because the center-bias for ROI prediction does not appear to significantly affect the results for the IQA, unlike for the general saliency prediction. In addition, it is reasonable to argue that the slight differences in the results may be due to the variance arising from the randomnesss at the network initialization/learning stage, or it may be due to the varying influence of center-bias effect according to the underlying nature of each dataset.
\\~\\
\textbf{Effect of linear normalization}
In order to test the effectiveness of linear normalization (Eq. 2), we re-performed the experiment by using the commonly used softmax normalization (Eq. 1), instead; please refer to the results under the ‘with softmax normalization’ row of Table \ref{table:supplement}. The experimental results indicate that the performance when using linear normalization is better than when using softmax normalization, and these findings are in agreement with the results of \cite{yang2019dilated} supporting that linear normalization better reflects the human eye fixation phenomenon.


\end{document}